\documentstyle[12pt]{article}
\begin{document}
\thispagestyle{empty}
\begin{center}
\LARGE\tt\bf{CMBR in G\"{o}del Universe with torsion:A possible test to Einstein-Cartan Gravity?}
\end{center}
\vspace{2.5cm}
\begin{center}{\large L.C. Garcia de Andrade\footnote{Departamento de F\'{\i}sica Teorica-UERJ.
Rua S\~{a}o Fco. Xavier 524, Rio de Janeiro, RJ
Maracan\~{a}, CEP:20550-003 , Brasil.
E-Mail.: garcia@dft.if.uerj.br}}
\end{center}
\vspace{2.0cm}
\begin{abstract}
A rotating universe represented by the G\"{o}del metric in spacetimes with Cartan  torsion is investigated where the Cosmic Microwave Background
Radiation (CMBR) is computed from the G\"{o}del rotation of the universe and the spin density of the spinning fluid in Einstein-Cartan gravity.The autoparallel equation in Riemann-Cartan  spacetime is shown to lead to the evolution equation of the cosmological perturbation  where the spin-rotation coupling is the source of the growth of inhomogeneities.It is shown that temperature anisotropy is within the limits of COBE constraint yielding a value of $\frac{{\delta}T}{T}=10^{-7}$.This result opens the possibility of testing Einstein-Cartan gravity from a cosmological experiment such as PLANCK or MAP. 
\end{abstract}
\vspace{2.0cm}
\newpage
Many years ago J.Silk \cite{1} has discussed the stability of G\"{o}del universe in the realm of general relativistic cosmology.
Among other results he showed that the rotating universe is stable along the to perturbations in the plane of rotation while is unstable along the rotation axis;and that the CMBR temperature anisotropy depends  on the G\"{o}del rotation.In this letter we show that Silk result can be extended to Riemann-Cartan \cite{2}  spacetimes with torsion where use is made of the autoparallels \cite{3}instead of geodesics in a manner that we are led to the evolution of the G\"{o}del perturbation of densities in the spinning fluid in Einstein-Cartan gravity (EC).The idea of introducing the spin-torsion coupling with the G\"{o}del rotation leads to a source coupling term between the spin and rotation.Solution of this equation allow us to investigate the stability of the G\"{o}del universe with torsion.The evolution of  the density perturbations allow us  to write an expression for the anisotropy in the CMBR as done in GR.Here we are able to compute the temperature anisotropy taking into account the spin-rotation effect induced by torsion and show that this is well within the COBE  satellite constraints.Let us now consider the G\"{o}del metric in the form \cite{4}
\begin{equation}
ds^{2}=a^{2}[(dx^{0})^{2}-(dx^{1})^{2}+\frac{e^{2x^{1}}}{2}(dx^{2})^{2}-(dx^{3})^{2}+2e^{x^{1}}dx^{0}dx^{2}]
\label{1}
\end{equation}
where $a^{2}$ is the the G\"{o}del scale parameter and ${\Omega}=(0,0,\frac{c}{{\alpha}\sqrt{2}})$.The autoparallel equation is given 
\begin{equation}
\frac{d}{ds}v^{\mu}+[-{\Gamma}^{\mu}_{{\alpha}{\beta}}+2Q^{\mu}_({{\alpha}{\beta}})]v^{\alpha} v^{\beta}=0
\label{2}
\end{equation}
where ${\Gamma}^{\mu}_{{\alpha}{\beta}}$ is the Riemannian connection or the Christoffel connection and $i=1,2,3$ and ${\mu}=0,1,2,3$ and the torsion tensor $Q^{\mu}_{{\alpha}{\beta}}=S^{\mu}_{\alpha}v_{\beta}$  is splitted in terms of the spin density tensor according to the Frenkel condition $S^{\mu}_{\alpha}v^{\alpha}=0$ where the spin density tensor $S_{{\mu}{\alpha}}=-S_{{\alpha}{\mu}}$.Taking the divergence of equation
\cite{2} yields  
\begin{equation}
\frac{d}{ds}{{\partial}_{\mu}v^{\mu}}+ [-{\Gamma}^{\mu}_{{\alpha}{\beta}}+2Q^{\mu}_({{\alpha}{\beta}})]v^{\alpha} v^{\beta}+2[-{\partial}{\Gamma}^{\mu}_{{\alpha}{\beta}}+2{\partial}_{\mu}Q^{\mu}_({{\alpha}{\beta}})]({\partial}_{\mu}v^{\alpha}) v^{\beta}=0
\label{3}
\end{equation}
From the conservation equation in Riemann-Cartan spacetime 
\begin{equation}
{\nabla}_{\mu}({\rho}v^{\mu})=0
\label{4}
\end{equation}
where ${\nabla}_{\mu}$ is the Riemann-Cartan covariant derivative.This expression substituted into the autoparallel equation yields
\begin{equation}
a^{2}\frac{{\partial}^{2}}{{{\partial}t}^{2}}ln{\rho}+\frac{c^{2}}{2}g^{{\lambda}{\alpha}}h_{{00},{\alpha}{\lambda}}=c\frac{{\partial}{e^{x^{1}}v^{2}}}{{\partial}x^{1}}-2ce^{-x^{1}}\frac{{\partial}v^{1}}{{\partial}x^{2}}-2[a^{2}{\partial}_{\mu}Q^{\mu}_{{\alpha}{\beta}}v^{\alpha}v^{\beta}+2 Q^{\mu}_{{\alpha}{\beta}}{\omega}^{\alpha}_{\mu}v^{\beta}]
\label{5}
\end{equation}
This equation reduces to the Silk geodesic equation \cite{1} in G\"{o}del spacetime when torsion $Q^{\mu}_{{\alpha}{\beta}}$ vanishes.Here we also used the definition of the rotation tensor \cite{5} as ${\omega}_{{\alpha}{\beta}}={\partial}_{[{\alpha}}v_{\beta}]$.The spinning fluid in EC gravity discussed above along with the Silk relation \cite{1}
\begin{equation}
g^{{\lambda}{\alpha}}h_{{00},{\lambda}{\alpha}}=\frac{8{\pi}G}{c^{2}}(a^{2}{\delta}{\rho}+\frac{3c^{2}}{8{\pi}a^{2}}h_{00})
\label{6}
\end{equation}
substitution of the equation (\ref{6}) into equation (\ref{5}) one obtains the evolution of the cosmological density perturbations for the G\"{o}del model  
\begin{equation}
\frac{d^{2}}{dt^{2}}{\delta}-4{\pi}G{\rho}\frac{{k_{3}}^{2}}{k^{2}+3}{\delta}=2[a^{2}{\partial}_{\mu}Q^{\mu}_{{\alpha}{\beta}}v^{\alpha}v^{\beta}+2 \frac{Q^{\mu}_{{\alpha}{\beta}}}{a^{2}}{\omega}^{\alpha}_{\mu}v^{\beta}]
\label{7}
\end{equation}
where we have used the spatial Fourier decomposition for the density contrast ${\delta}=\frac{{\delta}{\rho}}{\rho}=exp(k_{\lambda}x^{\lambda})$.Here to simplify matters which do not violate any experimental results one considers the approximation $|\frac{Q^{\mu}_{{\alpha}{\beta}}}{a^{2}}|<<1$ which allow  us to drop the second term on the RHS of equation (\ref{7}) and reduces this equation to
\begin{equation}
\frac{d^{2}}{dt^{2}}{\delta}-4{\pi}G{\rho}\frac{{k_{3}}^{2}}{k^{2}+3}{\delta}=2[a^{2}({\partial}_{\mu}S^{\mu}_{\alpha})v^{\alpha}+2 {S^{\mu}_{\alpha}}{\omega}_{{\mu}{\alpha}}v^{\beta}v^{\alpha}]
\label{8}  
\end{equation}
From the Frenkel condition one obtains
\begin{equation}
\frac{d^{2}}{dt^{2}}{\delta}-4{\pi}G{\rho}\frac{{k_{3}}^{2}}{k^{2}+3}{\delta}=2[a^{2}({\partial}_{\mu}S^{\mu}_{\alpha})v^{\alpha}]
\label{9}
\end{equation}
Since the Frenkel condition of the Weyssenhoff spinning fluid yields
\begin{equation}
{\partial}_{\mu}(S^{\mu}_{\alpha}v^{\alpha})=({\partial}_{\mu}S^{\mu}_{\alpha})v^{\alpha}+S^{{\mu}{\alpha}}{\omega}_{{\mu}{\alpha}}
\label{10}
\end{equation}
Substution of (\ref{10}) and (\ref{9}) yields 
\begin{equation}
\frac{d^{2}}{dt^{2}}{\delta}-4{\pi}G{\rho}\frac{{k_{3}}^{2}}{k^{2}+3}{\delta}+{\Omega}S=0
\label{11}
\end{equation}
where we have used $k^{2}=k^{\lambda}k_{\lambda}$, ${\Omega}={\omega}_{12}$ and $S=S^{12}$ is the only nonvanishing component of the spin density tensor.Here the last term on the LHS of expression (\ref{11}) represents the spin-rotation coupling induced by Cartan torsion.By solving the evolution equation of the G\"{o}del perturbation equation (\ref{11}) one obtains
\begin{equation}
{\delta}=sinh{At}-\frac{{\Omega}S}{A}
\label{12}
\end{equation}
where  $A=4{\pi}G{\rho}\frac{{k_{3}}^{2}}{k^{2}+3}$.Since ${\delta}=\frac{{\delta}T}{T}$ this allow us to place a limit on the anisotropy in CMBR computing this term.To achieve this aim we need to obtain the G\"{o}del rotation ${\Omega}$ and the spin density S.But this is already computed.The spin density of the Universe is given in the book of de Sabbata and Sivaram on spin and torsion in gravitation \cite{3} which is $S=10^{-27}gcm^{-2}s^{-1}$ and the rotation of the Universe is given in the Silk paper as the minimum value to be important for galaxy formation which is ${\Omega}=10^{-17} rad s^{-1}$ which by substitution into expression for the temperature anisotropy yields
\begin{equation}
|\frac{{\delta}T}{T}|= \frac{{\Omega}S}{4{\pi}G{\rho}}=10^{-7}
\label{13}
\end{equation}
where we have used the value of $10^{-30}g.cm^{-3}$ for the energy density of the universe.This result is within the COBE constraint of ${\delta}<10^{-5}$ and opens a possibility for testing EC gravity theory on a cosmological basis even by making use of device that measure CMBR such as the PLANCK or MAP modern experiments.Therefore as long as this model is concerned Einstein-Cartan cosmology seems to have some hope to be tested from any astronomical measurements concerning CMB.Recently \cite{6} we have obtained limits of the order ${\delta}=10^{-45}$ by investigating the origin of galaxy rotation using a spinning model in the EC gravity, which are far beyond quantum limitations of any device of modern astronomy observations.Note that here we do not make use torsion in the very early universe where the spin-torsion densities are higher.Besides recently we have found that through the study of CMBR is possible to find some Doppler effect \cite{7} that vanishes in general relativity but do not vanish in EC gravity.This opens new possibilities of detecting torsion by generalizing the Sachs-Wolfe effect.Topological defects such as superconducting cosmic strings \cite{8} with torsion also opens another venue to investigate electromagnetic radiation of some clusters and places limits on torsion from COBE data. 

\section*{Acknowledgements}
I am very much indebt to Professors F.W.Hehl,T.Villela Neto,A.Wuensche,A.Ribeiro,and O.Aguiar for helpful discussions on the subject of this paper.Thanks are also due to CNPq. (Brazilian Government Agency) and to FAPESP for financial support.


\begin{thebibliography}{8}
\bibitem{1}J.Silk,Mont.Not.Roy.Astr.Soc.(1970)147,13.
\bibitem{2}V.de Sabbata and C.Sivaram,Spin and Torsion in Gravitation,(1994),World Scientific.
\bibitem{3}L.C.Garcia de Andrade,On non-Riemannian domain walls,General Relativity and Gravitation Journal,(1998),30,11,1629. 
\bibitem{4}K.G\"{o}del,Rev.Mod.Phys.(1949).
\bibitem{5}M.Novello and M.Reboucas,ApJ,225,(1978),719.  
\bibitem{6}L.C.Garcia de Andrade and A.Ribeiro,Origin of angular momentum of galaxies in the context of a spinning fluid model in Einstein-Cartan gravity,(2000),Preprint-Departamento de Matematica Aplicada-Unicamp. 
\bibitem{7}L.C.Garcia de Andrade,Signature of Torsion in CMBR?,Preprint DFT-UERJ-(2000). 
\bibitem{8}C.N.Ferreira,H.Mosquera and L.C.Garcia de Andrade,Superconducting cosmic strings and torsion propagation,(2000),CBPF-preprint. 
\end{thebibliography}
\end{document}